\documentclass[twocolumn,showpacs,preprintnumbers,amsmath,amssymb]{revtex4}
\usepackage{graphicx}
\usepackage{dcolumn}
\usepackage{bm}
\def\bea {\begin{eqnarray}}
\def\eea {\end{eqnarray}}

\def\be {\begin{equation}}
\def\ee {\end{equation}}

\begin{document}
\title{Do we see change of phase in proton-proton collisions at the Large Hadron Collider?}
\author{Premomoy Ghosh}
\email{prem@vecc.gov.in}
\author{Sanjib Muhuri}
\email{sanjibmuhuri@vecc.gov.in}
\address{Variable Energy Cyclotron Centre, Kolkata 700 064, India}                 

\date{\today}
\begin{abstract}
High multiplicity events in proton-proton collisions at the Large Hadron Collider exhibit features resembling those found in relativistic nuclear 
collisions, indicating possibility of formation of similar partonic medium. Formation of partonic medium in hadronic collision means happening 
of change in phase of matter, as is predicted by quantum chromodynamics. Analyzing the identified particle spectra data from $proton-proton$ collision, 
we observe a signal of plausible change of phase between thermalized partonic matter and hadronic one, in terms of change in effective number 
of degrees of freedom, as has been revealed by the latest calculation in the framework of the lattice quantum chromodynamics. The result is compared 
with similar analysis of $proton-lead$ data and simulated events of $proton-proton$ collisions. 
 
\end{abstract}
\maketitle

\section{Introduction}
\label{}
The minuscular universe of high energy-density and temperature, immediately after the Big Bang, evolved through a state of an exotic phase of 
matter, the Quark-Gluon Plasma (QGP) \cite{ref01}, that survived for some microseconds only. The QGP - the plasma state in quantum chromodynamics 
(QCD)\cite{ref02}, the theory of strong interaction - can be created also in the laboratory \cite{ref03, ref04, ref05, ref06}, in relativistic heavy-ion collisions. 
The formation of the QCD plasma necessitates accomplishment of thermodynamic equilibrium. In heavy-ion collisions,  the local thermodynamic equilibrium 
is assumed to obtain the equation of state for the space-time evolution of the system in the framework of the relativistic hydrodynamics \cite{ref07}. Eventually, 
the formation of QGP in heavy-ion collisions is established mainly by characterizing the system in the hydrodynamic model, implying occurrence of the QCD 
phase-change. The proton-proton ($pp$) collision, on the other hand, is considered as the elementary interaction by most of the topical models which do not 
endorse formation of any medium, whatsoever. One may also question if the mean free path of the constituent particles, the size and the lifetime of the 
produced medium, if any, in $pp$ collisions are conducive to the formation of QGP or not. Nevertheless, the recent data of $pp$ collisions, particularly for 
the high-multiplicity events, at the Large Hadron Collider (LHC) at CERN show certain features \cite{ref08, ref09, ref10}, which are typical of the hydrodynamic 
medium formed in the relativistic heavy ion collisions. Several of subsequent phenomenological studies with the LHC $pp$ data indicate the formation 
\cite{ref11, ref12, ref13} of collective medium. Further, while some theoretical approaches\cite{ref14, ref15, ref16} support applicability of relativistic 
hydrodynamics in high multiplicity $pp$ collisions, formation of a non-hadronic medium has been shown \cite{ref14} to be energetically possible in $pp$ 
collisions at the LHC. A study \cite{ref17} in search for the change in phase in $pp$ collisions, analyzing a wide range of centre-of-mass energy, includes 
the LHC data also. The prima facie observations from all these studies in $pp$ collisions at the LHC are quite encouraging for one to follow the approach 
adopted in search for the QGP in heavy-ion collisions - assume local thermal equilibrium and search for signals for formation of QGP or equivalently the 
QCD phase-change . We analyze the data of $pp$ collisions at the LHC to search for a signal of change of phase, in QCD, from thermalized partonic 
matter to hadronic one, in terms of change in effective number of degrees of freedom of the system formed in the collision, as a function of temperature.
The signal has been predicted \cite{ref18, ref19, ref20, ref21, ref22, ref23} by the Lattice Quantum Chromodynamics (LQCD), the well accepted tool for 
non-perturbative calculations in QCD. 

\section{The signal for change of phase}
\label{}
The LQCD - predicted signal for the phase transition or for the crossover in high energy heavy-ion and $pp$ collisions calls for a 
fast rise of $\epsilon / T^{4}$ or $\sigma / T^{3}$ (where $T$ is temperature, $\epsilon$ is the energy-density and $\sigma$ is the entropy-density of the 
system) over a small change in temperature, reflecting rapid increase in the effective number of degrees of freedom indicating 
a change in phase. While a sharp, step-like increase manifests the first order phase transition, a smooth but rapid rise indicates either the crossover or a 
second order phase transition. The lattice QCD simulation is continuously progressing \cite{ref24} towards more realistic calculations, in terms of number of quark flavors 
and their masses, minimization of lattice spacings etc. The latest LQCD calculations result in a crossover\cite{ref23} in the QCD transition. At temperature
lower than the critical region of rapid rise, the system is populated mainly by pions, forming a hadron gas. In this temperature range, however, the lattice calculation 
of $\epsilon$ and so $\sigma$ is difficult as they are exponentially suppressed\cite{ref21}. At temperature higher than the critical region, where the ideal gas 
limit of degrees of freedom is reached asymptotically, the system corresponds to ideal gas consisting of partons. In absence of a straightforward method of 
estimation of temperature from data, the predicted behavior of temperature dependence of  $\epsilon / T^{4}$ and $\sigma / T^{3}$ as a signal for the phase 
transition or the crossover is still a far from convincing experimental verification even for the heavy-ion collisions. Adding to the complexity, in the real-life 
scenario of experiments, the finite size effect\cite{ref25, ref26} of the system formed in the collision, may cause widening of the width of transition or 
smoothening of singularities in the LQCD-predicted dependence of the degrees of freedom. 

\section{Methodology}
\label{}
The experimental search for the phase transition or for the crossover in the data of high energy collisions involves connecting thermodynamical variables 
with measurable observables. The rapidity density, $dN_{ch}/dy$, (where $N_{ch}$ = the number of charged hadrons, the rapidity, $y = 
(1/2) \ln\frac{E + p_{L}}{E - p_{L}}$, $E$ and $p_{L}$ are respectively the energy and longitudinal component of momentum of a particle) or the pseudorapidity
density, $dN_{ch}/d\eta$, (where the pseudorapidity, $\eta = -ln[tan(\theta/2)]$ and  $\theta$ is the polar angle of the particle with respect to the beam direction) 
and the transverse momentum, $p_{T}$, of produced charged hadrons are primary measurable observables in high energy experiments 
which are connected to thermodynamic variables involved in the LQCD-predicted signal for the change in phase.

According to the relativistic hydrodynamics, the rapidity or the pseudorapidity density reflects the entropy-density ($\sigma$) created early in the collision. The $\sigma$ 
depends only on the initial energy-density ($\epsilon$) and remains constant throughout the evolution of the system. It is worth mentioning that even for 
$d\langle N_{ch} \rangle /d\eta \approx 6$, the $\epsilon$ has been shown\cite{ref14} to be sufficient for formation of a non-hadronic medium in $pp$ collisions at 
$\sqrt {s}$ = 7 TeV. The Compact Muon Solenoid (CMS) experiment has recorded \cite{ref27, ref28} sufficient statistics of high multiplicity $pp$ events with 
$d\langle N_{ch} \rangle /d\eta$ going up to $\approx 30$, providing a wide range of data to look for effects of variation in observables involving energy-density. 
As the objective of this study does not include evaluation of the absolute value of the energy-density, the following simple formalism  \cite {ref19} as proposed by 
Bjorken could be sufficient for estimation of the energy-density: 

\begin{equation}
\epsilon_{BJ} \simeq \frac{\frac{dN_{ch}}{d\eta}. \frac {3}{2}{\langle p_{T} \rangle}}{V}       
\end{equation} 

where, $V$ is the effective volume occupied by the particles and is given by  $S. c\tau $, $S$ being the interaction area and $c.\tau$, a longitudinal 
dimension, considered to be about 1$fm$ in Bjorken model. The estimate of entropy density, therefore, comes as:
\begin{equation}
\sigma  \simeq \frac{\frac{dN_{ch}}{d\eta}. \frac {3}{2}}{S}     
\end{equation}  
where $\sigma = \epsilon / \langle p_{T} \rangle$ and  $S = \pi R^2$, $R$ is the radius of the interaction area.

\begin{center}
\begin{figure}
\includegraphics[scale=0.40]{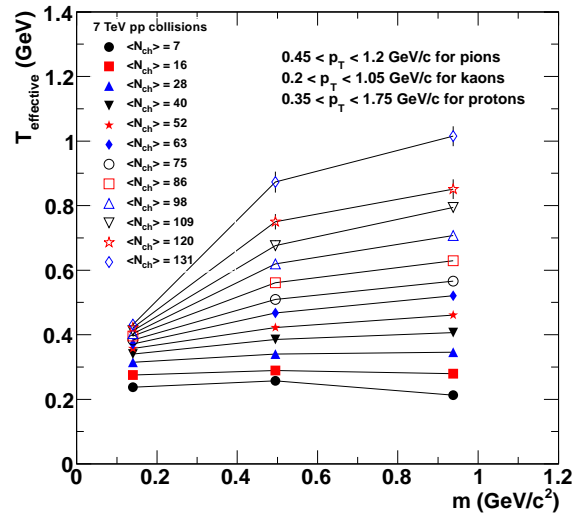}
\caption{The inverse slope parameter $T_{effective}$ as a function of mass of identified particles ($\pi^\pm, K^\pm$  and $p(\bar p)$), obtained from the 
measured transverse momentum spectra from non-single diffractive events of $pp$ collisions as measured by the CMS experiment  \cite {ref27,ref28} at 
$\sqrt {s}$ = 7 TeV. $\langle N_{ch} \rangle$ is the mean multiplicity of charged particles. The solid lines drawn joining the data points are to guide the eye.}
\label{fig:massordering} 
\end{figure}
\end{center}

The low $p_{T}$ part of the $p_{T}$ - spectra of produced particles from high energy collisions contain information on temperature as well as on transverse 
expansion of the system. As disentangling the effect of the temperature and the contribution from the transverse expansion in the $p_{T}$ - spectra is not 
possible, instead of exclusive measure of temperature, experimental analysis deal with parameters that reflect the temperature of the system. Either the 
mean transverse momentum, ($\langle p_{T} \rangle$), as proposed \cite{ref18} by Van Hove, or the slope of the transverse mass  ($m_{T}$) - spectra, 
obtained from the $p_{T}$ - spectra (for a particle of mass $m$, $m_{T} = (m^2 + p_{T}^2)^{1/2}$), can be used for comparing thermal states of the system. 
The investigation of the LQCD predicted signal for change in phase in terms of effective number of degrees of freedom in $pp$ collisions has been 
studied \cite{ref17} in terms of the $\langle p_{T} \rangle$ dependence of $\sigma / {\langle p_{T} \rangle}^{3}$.  On the other hand, the $m_{T}$ - spectra 
corresponding to low - $p_{T}$ particles are usually satisfactorily fitted with exponential function of the form, ${dN}/{m_{T}dm_{T}} = C.exp(- {{m_{T}}/{T_{effective}}})$, 
where $T_{effective}$ is known as the inverse slope parameter. Increase in inverse slope parameter, $T_{effective}$ with mass, $m$, for the most commonly measured 
identified particles ($\pi^\pm$ , $K^\pm$, $p$ and $\bar p$), as has been observed \cite{ref29,ref30} in heavy-ion collisions as well as in the recent $pPb$ 
collisions \cite{ref31} at CERN is a well known phenomenon, attributed to the collective transverse flow of the medium formed in the collision. 

\section{Analysis $\&$ Results}
\label{}

\subsection{$pp$ data}
\label{}

The CMS experiment at the LHC has measured $p_{T}$-spectra \cite{ref27, ref28} of pions ($\pi^\pm$), kaons ($K^\pm$), and protons ($p$ and $\bar p$) over the 
rapidity range $|y|<1$ for the $pp$ collisions at $\sqrt {s}$ = 0.9, 2.76 and 7 TeV for several event classes selected on the basis of mean number of charged 
particles, $\langle N_{ch} \rangle$ in the pseudo-rapidity interval, $|\eta| <2.4$ or, in other words, as a function of  pseudorapidity density, 
$d\langle N_{ch} \rangle /d\eta$, reflecting "centrality" in $pp$ collisions. The measured $p_{T}$ - ranges are  (0.1 to 1.2) GeV/c for $\pi^\pm$, (0.2 to 1.050) 
GeV/c for $K^\pm$ and (0.35 - 1.7) GeV/c for $p$ and $\bar p$.  In fact, by measuring the $d\langle N_{ch} \rangle /d\eta$ - dependent identified particle
spectra, the CMS experiment provides the unique opportunity to study the LQCD-predicted signal on change in phase in terms of the change in degrees of freedom
in $pp$ collisions at the LHC energies. It is worth mentioning that the same set of data, on analysis \cite {ref13} in the hydrodynamics 
motivated Boltzman-Gibbs blast-wave model \cite{ref32}, revealed collective transverse flow for the high multiplicity events. To check if the previously observed 
indication of the formation of collective medium in $pp$ collisions is revealed in the statistical model also, we look for the mass dependence of inverse slope 
parameter by fitting the $m_{T}$ - spectra (with the $\chi^2$ values in the range 0.3-2.48 for pions, 0.26-1.08 for kaons and 0.06-4.92 for protons) of identified 
particles obtained from the $p_{T}$ - spectra data \cite{ref28} at $\sqrt {s}$ = 7 TeV measured by the CMS experiment. To avoid the effect of resonance decays, 
the pions with $p_{T}$ \textless 0.45 GeV/c were excluded from the fitting-range. The increase of $T_{effective}$ with mass of identified particles for event classes 
of high $\langle N_{ch} \rangle$ reiterates (Figure~\ref{fig:massordering}) the finding of collective medium in high-multiplicity $pp$ collisions. The collectivity, 
however, does not confirm formation of the QGP or the QCD phase-change. 

\begin{center}
\begin{figure}
\includegraphics[scale=0.40]{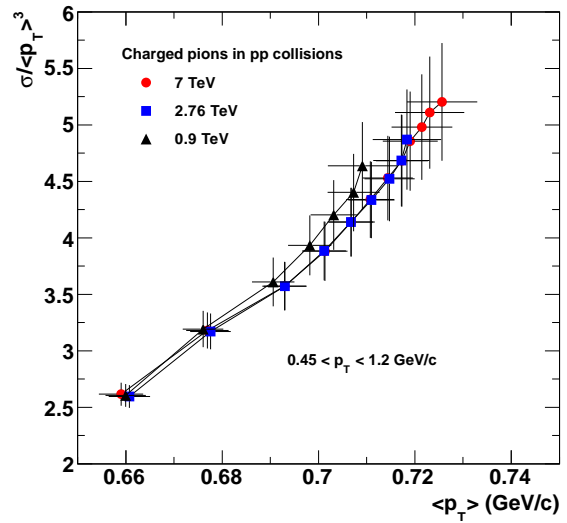}
\caption{The $\sigma / {\langle p_{T} \rangle}^{3}$, the effective number of degrees of freedom, as a function of the transverse momentum, $\langle p_{T} \rangle$, 
of pions as obtained from the measured \cite{ref28} spectra data. The solid lines drawn joining the data points are to guide the eye. The error bars include both the 
statistical and systematic uncertainties.}
\label{fig:sigmabyptcube_pi} 
\end{figure}
\end{center}

\begin{center}
\begin{figure}
\includegraphics[scale=0.40]{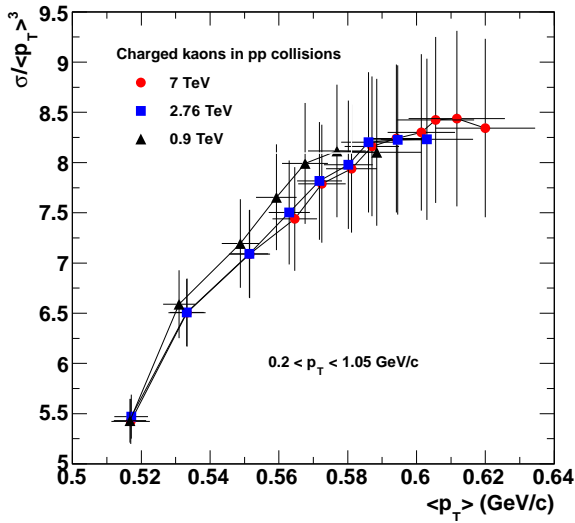}
\caption{The $\sigma / {\langle p_{T} \rangle}^{3}$, the effective number of degrees of freedom, as a function of the mean transverse momentum, $\langle p_{T} \rangle$, 
of kaons as obtained from the measured \cite{ref28} spectra data. The solid lines drawn joining the data points are to guide the eye. The error bars include both the 
statistical and systematic uncertainties.}
\label{fig:sigmabyptcube_K} 
\end{figure}
\end{center}

To characterize the collective medium so formed, we study  the $\langle p_{T} \rangle$ -  dependence of $\sigma / {\langle p_{T} \rangle}^{3}$, the effective number 
of degrees of freedom of the system formed in the collisions. In the study, we also take into consideration some model dependent estimations \cite{ref33, ref34} of 
the effect due to the transverse expansion on $\langle p_{T} \rangle$ to facilitate the study and the interpretation thereof. A model calculation \cite{ref33} 
of $\langle p_{T} \rangle$ as a function of the temperature for various hadron species show that the change in the shape of the $p_{T}$ - spectra or in the value of  
$\langle p_{T} \rangle$ due to the transverse expansion is dependent on the individual hadron species of different mass - while the protons gain most in 
$\langle p_{T} \rangle$, the pions lose some $\langle p_{T} \rangle$. According to another model calculation \cite{ref34}, the protons gain additional 
$\langle p_{T} \rangle$ as the protons decouple from the fireball after pions, experiencing larger flow-effect. Being weakly affected \cite{ref35} by re-scattering 
of hadrons and resonance decay, kaons are better observable in the search for change in phase. We, therefore, argue that instead of the $\langle p_{T} \rangle$ 
of several particles together, containing cumulative effect due to the transverse expansion of varying magnitude, the $\langle p_{T} \rangle$ of individual species 
may be more meaningful in reflecting temperature related characteristics of the system. We study the dependence of $\sigma / {\langle p_{T} \rangle}^{3}$ on the 
$\langle p_{T} \rangle$ for charged pions and charged kaons from the $pp$ collisions at $\sqrt {s}$ = 0.9, 2.76 and 7 TeV, separately, as kaons have been suggested 
\cite{ref35} to be a better observable particles and the charged pions, constituting the majority of the produced charged particles, grossly dominate the  average 
behavior of all charged particles. To calculate the entropy density, $\sigma$ following Bjorken's formalism \cite{ref19}, the radius of interaction area, $R$, is 
estimated from the HBT- radius \cite{ref36} that has dependence on the pair transverse momentum ($k_{T}$) and multiplicity. For a given $d{\langle N_{ch} \rangle}/d\eta$, 
high $k_{T}$ corresponds to the early stage of the collision, while low $k_{T}$ relates the final stage, close to the freeze-out, and includes the effect of particle 
activities in the transverse direction during the expansion. We use the radii for the highest  $k_{T}$ - range (0.5 to 1.0 GeV/c), as measured \cite{ref10} by the CMS 
experiment, to parameterize the radius of interaction as a function of $\langle N_{ch} \rangle$, as $R (\langle N_{ch} \rangle) = a. \langle N_{ch} \rangle^{1/3}$, 
where $a = 0.504 \pm 0.0097$ fm.  

Both the pions and the kaons show more or less smooth but rapid rise in $\sigma / {\langle p_{T} \rangle}^{3}$, indicating release of new degrees of freedom, over a 
small change of $\langle p_{T} \rangle$ in the lower $\langle p_{T} \rangle$ - range. Significantly, however, the dependence of $\sigma / {\langle p_{T} \rangle}^{3}$ 
at high $\langle p_{T} \rangle$ is different for pions and kaons. A change of phase cannot be concluded from the plots in Figure~\ref{fig:sigmabyptcube_pi} for the 
pions, because of the missing saturation effect at high $\langle p_{T} \rangle$ - range. For kaons at high $\langle p_{T} \rangle$, reflecting high temperature, 
the saturation in $\sigma / {\langle p_{T} \rangle}^{3}$, as depicted in Figure~\ref{fig:sigmabyptcube_K}, indicates that the highest possible number of degrees of freedom, 
corresponding to the partonic phase, is reached. Different responses by pions and kaons is rather consistent with previous studies \cite{ref33, ref34, ref35}. The loss 
of $\langle p_{T} \rangle$ for pions \cite{ref33} could be the reason for the disappearance of the saturation effect of $\sigma / {\langle p_{T} \rangle}^{3}$ at high 
$p_{T}$, corresponding to the high multiplicity $pp$ events. Also, the differential freeze-out of hadron species as a consequence of different elastic 
cross-sections \cite{ref34, ref37,ref38} may lead to the early kaon-freeze-out as compared to the pion-freeze-out and may provide an alternate explanation to the 
observed difference in response by the pions and the kaons. In case of early freeze-out, the kaons are likely to be less affected by the flow, that develops with time. 
The shape of the kaon $p_{T}$ -spectrum, therefore, may largely retain the effect of the temperature of the system formed in high energy collisions. One needs 
to note that, as the $\langle p_{T} \rangle$ only reflects the $T$ and does not give absolute measure of $T$, the $\sigma / {\langle p_{T} \rangle}^{3}$ also 
reflects the effective number of degrees of freedom and not necessarily quantitatively match $\sigma / T^{3}$. Nevertheless, the dependence of 
$\sigma / {\langle p_{T} \rangle}^{3}$ on $\langle p_{T} \rangle$ for the kaons reveals the qualitative feature of the LQCD-predicted signal for the change in phase 
of matter formed in $pp$ collisions in terms of change from high to low degrees of freedom with decreasing $\langle p_{T} \rangle$, the temperature-related observable. 
Importantly, the nature of the experimental finding of the change in phase matches the results of the latest LQCD calculations predicting a crossover \cite{ref23}.

\subsection{Simulation}
\label{}
\begin{center}
\begin{figure}
\includegraphics[scale=0.40]{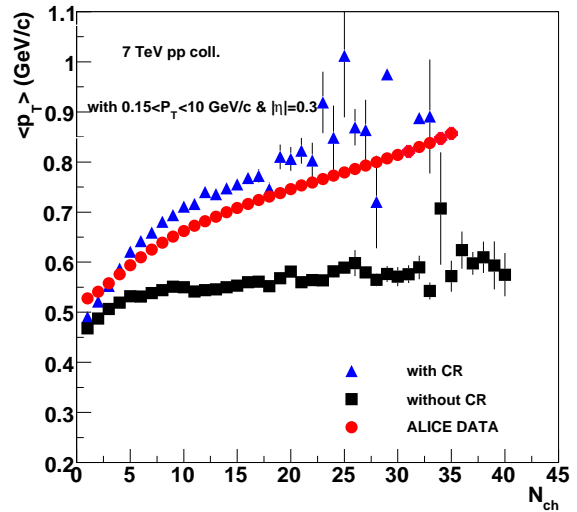}
\caption{Average transverse momentum, $\langle p_{T} \rangle$, as a function of number of charged-particle multiplicity, $N_{ch}$, as measured by ALICE \cite {ref39} 
is compared with the simulated events using PYTHIA8 Tune 4C generators. The large statistical uncertainties correspond to only ten thousand simulated 
events of each class.}
\label{fig:alice_setup} 
\end{figure}
\end{center}
\begin{center}
\begin{figure}
\includegraphics[scale=0.40]{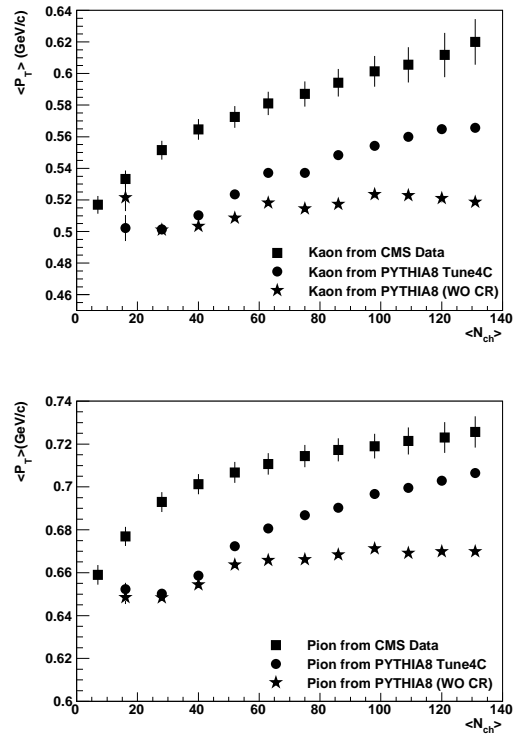}
\caption{Average transverse momentum, $\langle p_{T} \rangle$, as a function of mean charged-particle multiplicity, $\langle N_{ch} \rangle$, as measured by CMS \cite {ref27} 
is compared with the simulated events using PYTHIA8 Tune 4C generators with Color Reconnection on for $pp$ collisions at $\sqrt {s} =$ 7 TeV. The error bars include both the statistical and systematic uncertainties.}
\label{fig:meanptVsNch_piK} 
\end{figure}
\end{center}
\begin{center}
\begin{figure}
\includegraphics[scale=0.40]{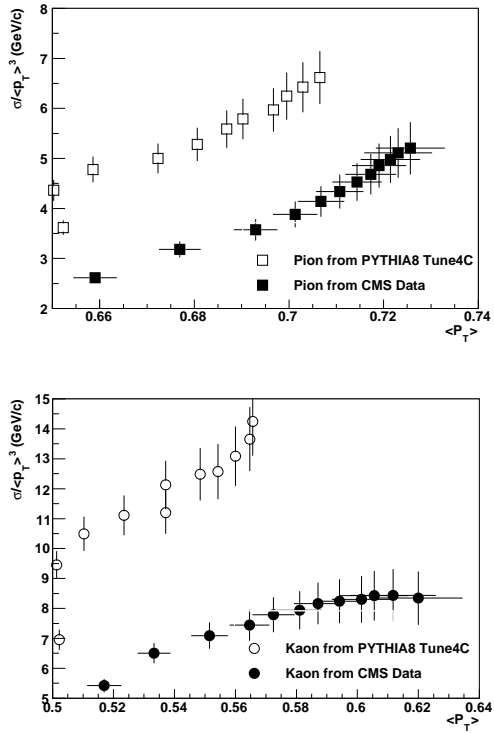}
\caption{The $\sigma / {\langle p_{T} \rangle}^{3}$, the effective number of degrees of freedom, as a function of the mean transverse momentum, $\langle p_{T} \rangle$, 
for pions and kaons as obtained from the measured \cite{ref31} spectra data and from the similar classes (depending on $dN_{ch}/d\eta$) of simulated events using PYTHIA8 
Tune 4C generators with Color Reconnection on for $pp$ collisions at $\sqrt {s} =$ 7 TeV. The error bars include both the statistical and systematic uncertainties.}
\label{fig:sigmabyptcube_simpiK} 
\end{figure}
\end{center}

As the analysis reveals a new feature in the $pp$ data, comparing the results with the conventional Monte Carlo Event Generators is virtually obligatory.
The standard event generators in use, however, have several versions tuned to explain different features of the LHC data. In that respect, we find the PYTHIA8.1 
Tune 4C, that, besides some other features, explains \cite{ref39} the $\langle p_{T} \rangle$ - dependence on $\langle N_{ch} \rangle$ in $pp$ collisions as measured 
by ALICE, to be relevant to compare the findings. It is important to mention that ALICE finds \cite{ref39} that the $\langle p_{T} \rangle$ - dependence on $\langle N_{ch} \rangle$ for 
charged hadrons in $pp$ collisions, in the $p_{T}$ - range up to 10 GeV/c,  is well described by PYTHIA8.1 Tune 4C by revoking Color Reconnection (CR), while generated 
events without CR do not reproduce the dependence. The CR phenomenon was introduced in the Multiple Partonic Interaction (MPI)  models to explain the increasing trend
of $\langle p_{T} \rangle$ as a function of produced charged particles in $pp$ or $(p \bar p)$ collisions. According to the MPI models, in high multiplicity events, the multiparticle
production results from a large number of MPIs. In CR mechanism, the final state partons from different MPIs get connected by color strings which follow the
movements of the partonic end-points, resulting a transverse - boost in the string fragmentation process. The effect is manifested in the observed dependence of 
$\langle p_{T} \rangle$ on $N_{ch}$. Particle production of independent MPIs is expected to result flat $\langle p_{T} \rangle$ as a function of $N_{ch}$. Using the PYTHIA8.1 
Tune 4C, we generate $pp$ events at $\sqrt {s} = $ 7 TeV, with both the options, CR-on and CR-off and plot in Figure~\ref{fig:alice_setup}, basically reproducing the ALICE 
plots. As the ALICE data of charged hadrons in the $p_{T}$ - range up to 10 GeV/c is consistent with the PYTHIA8.1 Tune 4C generated events, we compare the multiplicity
dependent $\langle p_{T} \rangle$ for the pions and the kaons spectra from the simulated events with the measured \cite{ref27} spectra that have been used in this work. We plot
$\langle p_{T} \rangle$ as a function of $\langle N_{ch} \rangle$ as obtained from the CMS data \cite{ref27, ref28} (within $|\eta| <2.4$, while the $p_{T}$ - ranges are  
(0.1 to 1.2) GeV/c for $\pi^\pm$, (0.2 to 1.050) GeV/c for $K^\pm$) and from the generated events with CR-on and CR-off. As it is clear from the Figure~\ref{fig:meanptVsNch_piK}, though the CR in the particular PYTHIA tune, in use, gives the boost in the $p_{T}$ of the produced particles, it is far from the measured ones in the given kinematic ranges, for all the 
event - classes depending on $\langle N_{ch} \rangle$. In spite of this mismatch, for the sake of completeness, we repeat our analysis related to the LQCD prediction on the 
simulated events with the CR-on. The obvious mismatch between the data and the simulated events with color reconnection is depicted in Figure~\ref{fig:sigmabyptcube_simpiK}. 
It is important to note at this point that the temperature of a thermalized partonic medium, if formed, should ideally be reflected by the low - $p_{T}$ particles (usually 
$p_{T} \textless$ 2 GeV/c, as has been considered in heavy-ion collisions) and the CMS data meet the criterion. 

\subsection{$pPb$ data}
\label{}

\begin{center}
\begin{figure}
\includegraphics[scale=0.40]{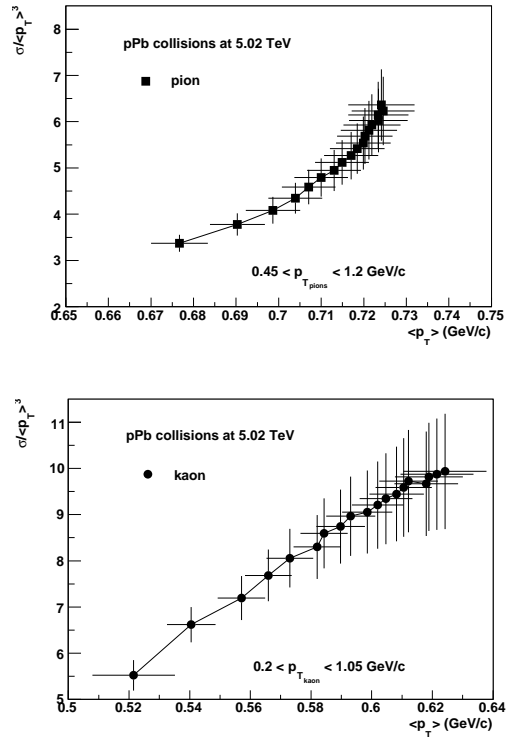}
\caption{The $\sigma / {\langle p_{T} \rangle}^{3}$, the effective number of degrees of freedom, as a function of the mean transverse momentum, $\langle p_{T} \rangle$, 
for pions and kaons as obtained from the measured \cite{ref31} spectra data from $pPb$ collisions at 5.02 TeV. The solid lines drawn joining the data points are to guide 
the eye. The error bars include both the statistical and systematic uncertainties.}
\label{fig:sigmabyptcube_piK_pPb} 
\end{figure}
\end{center}

The observation from the analysis of the identified spectra data of $pp$ collisions motivates us to analyze the identified spectra data of $proton-lead$ ($pPb$) 
collisions \cite {ref31} at $\sqrt s $ = 5.02 TeV following similar methodology, for comparison. We choose the $pPb$ data for comparison because of stronger 
evidence of collectivity  \cite{ref31, ref40, ref41, ref42} in the pPb collisions. Besides, availability of identified particle spectra data for different class of $pPb$ events
depending on the pseudorapidity density (similar classification as in $pp$ collisions) and remarkably similar \cite{ref43} $N_{ch}$ - dependence of radii in all $k_{T}$ 
bins for the $pp$ and $pPb$ collisions allow comparison on the same footings. As can be seen in the Figure~\ref{fig:sigmabyptcube_piK_pPb}, the $pPb$ data reveals
similar features as exhibited by the $pp$ data, though the effect of saturation of $\sigma / {\langle p_{T} \rangle}^{3}$ for kaons becomes weaker.

\section{Discussions and Conclusions}
\label{}
The search for thermalized partonic matter in the so-called elementary ($pp$ or $p \bar p$) interactions is not new. Prior to LHC also, there
have been several studies \cite {ref18, ref20, ref44, ref45, ref46}, addressing the de-confinement and formation of hydrodynamic medium in $p \bar p$ 
collisions. This study reveals an indication of possible change of phase in $pp$ collision data in terms of change in effective number of degrees of freedom 
of the system. On the issue of thermal equilibrium, one can assume that the measured final state particle multiplicity in high-multiplicity $pp$ events at LHC, being comparable  
\cite {ref15} with that in $CuCu$ collisions at $\sqrt {s_{NN}}$ = 200 GeV at RHIC, may be adequate enough to fulfill the condition. We emphasize that the observed 
dependence of the degrees of freedom on $\langle p_{T} \rangle$ at large values, equivalent to high temperature, corresponds to the high multiplicity $pp$ events. It 
is noteworthy that the liberation of effective number of degrees of freedom for a given pseudorapidity density and for a given $p_{T}$ - range, is independent of 
the centre-of-mass energy. Summarily, the analysis of the data of $pp$ collisions at the LHC, in terms of the nature of dependence of 
$\sigma / {\langle p_{T} \rangle}^{3}$, the effective number of degrees of freedom,  on $\langle p_{T} \rangle$, reflecting the temperature of the system, 
as measured from the $p_{T}$ - spectra of charged kaons, corroborates with the LQCD - calculation on change in phase. We repeat that though the QCD 
predicted change of phase has experimentally been proved \cite{ref03,ref04,ref05,ref06} in heavy-ion collisions by characterizing the medium formed after collisions,
the experimental verification of the LQCD - calculated signal for the change of phase in terms of change in degrees of freedom in still not convincingly established. 
This may be naively attributed to the dampening effect due the collective behavior of the large-size fireball formed in the heavy-ion collisions. That way, 
the $p_{T}$ - spectra of particles produced in a smaller fireball formed in $pp$ collision may appear to be a befitting system for the search of the 
LQCD - calculated signal. Also, the species-dependent response of the behavior of change in effective number of degrees of freedom, as has been revealed from 
this study, may become an useful tool in search for the LQCD - predicted signal in heavy-ion collisions. While the finding can not definitely be considered as the 
"smoking gun" on the change of phase in $pp$ collisions, it certainly reveals a new feature, significant enough in furthering the study in search of the collective medium
in high-multiplicity $pp$ events and in characterization of such medium at higher LHC energies, scheduled in the Run-II of the LHC in 2015.

\end{document}